\documentstyle[epic,eepic]{article}
\begin{document}
\title{Interferometry as a binary decision problem}
\author{Matteo G. A. Paris \\
Arbeitsgruppe 'Nichtklassiche Strahlung' der Max-Planck-Gesellschaft \\
         Rudower Chaussee 5, 12489 Berlin,Germany \\
         {\tt PARIS@PHOTON.FTA-BERLIN.DE}}
\date{}
\maketitle
\begin{abstract}
       Binary decision theory has been applied to the general
       interferometric problem. Optimal detection scheme---according
       to the Neyman-Pearson criterion---has been considered for
       different phase-enhanced states of radiation field, and the
       corresponding bounds on minimum detectable phase shift has
       been evaluated. A general bound on interferometric precision
       has been also obtained in terms of photon number fluctuations
       of the signal mode carrying the phase information.
\end{abstract}
\section{Introduction}
\label{s:int}
An interferometer is an optical device devised to reveal very small
variations in the optical path of a light beam. This is usually
accomplished by considering two parts of a quantum state traveling
along different routes, in a way they accumulate different amount
of phase. In such a general scheme the precision in measuring the phase
depends not only on the involved quantum state, but also on the specific
interferometric setup. In order to derive a general bound on the precision
of phase measurement a more abstract scheme has to be considered.
Here we will consider interferometry as a binary decision problem, where
the two signals are only differentiated by the occurrence of a phase shift.
Actually, this is more similar to a communication problem with the phase
shift playing the role of encoded information. Nonetheless, it appears
intuitively obvious that any form of conventional interferometry cannot
lead to better performance than this communication variety. \par
An abstract outline of an interferometric detection scheme is shown in
Fig. \ref{f:out}. An initially prepared state of radiation, say
$\hat\rho_0$, travels along the interferometer and it is eventually
measured by some detectors, denoted by $D$. The latter is described by
an operator-valued probability measure $d\hat\mu (x)$, being $x \in X$
the set of the possible detection outcomes. If some environmental parameter
changes then also the optical path is subjected to variation, thus leading a
phase-shift $\varphi$ on the signal mode. \par
The aim of the detection scheme $d\hat\mu (x)$ is that of discriminating
in between $\hat\rho_0$ and its phase-shifted version $\hat\rho_1=\exp\{i\hat
n \varphi\} \hat\rho_0 \exp\{-i\hat n \varphi\}$, which results if some
perturbations have been occurred. An optimized interferometer is able to
tell the $\hat\rho$'s apart for $\varphi$ smaller as possible. \par
This way of posing the interferometric problem naturally leads to view it
as a binary decision problem, to which results and methods from quantum
detection theory can be applied\cite{hel,hol}. Here, the phase-shift
$\varphi$ plays the role of a parameter, labeling one of the two possible
outputs from the interferometer, namely the perturbed state $\hat\rho_1$.
Indeed, this approach can be useful as it does not refer to any specific
detection scheme for the final stage of the interferometer. Thus,
ultimate quantum limit on the interferometric precision can be obtained
for specific classes of phase-enhanced states of radiation field
\footnote{\footnotesize Usually in optimizing interferometry just the
opposite route has been followed. After fixing some interferometric
setup precision has been optimized over the states of radiation
\cite{yur,scu,bur,rip}.}
. \par
\vspace{25pt}
\begin{figure}[thb]
\unitlength=1mm
\linethickness{0.4pt}
\begin{picture}(126.00,29.00)
\dottedline{1}(30,20)(32,17.5)(34,15.5)(36,14)(38,13)(40,12.1)(42,11.2)(44,10.5)(45.5,10)
\dottedline{1}(70,20)(68,17.5)(66,15.5)(64,14)(62,13)(60,12.1)(58,11.2)(56,10.5)(54.5,10)
\put(50,20){\vector(1,0){0.}}\put(36,13.75){\vector(1,-1){0.}}
\put(90.00,20.00){\shade{\arc{10.}{-1.57}{1.57}}}
\put(90.00,15.00){\line(0,1){10.00}}\put(64,14.25){\vector(1,1){0.}}
\put(10.00,20.00){\circle*{2.00}}\put(10.00,20.00){\line(1,0){80.00}}
\put(95.00,20.00){\line(1,0){15.00}}\put(110.00,20.00){\circle*{2.00}}
\put(50.00,10.00){\circle{9.0}}\put(125.00,28.00){\circle*{2.00}}
\put(130.00,28.00){${\cal H}_0 $}\put(125.00,12.00){\circle*{2.00}}
\put(130.00,12.00){${\cal H}_1$}
\put(10.00,10.00){\makebox(0,0)[cc]{\large  $\hat\rho_0$}}
\put(85.00,10.00){\makebox(0,0)[cc]{\large$d\hat\mu (x)$}}
\put(97.00,27.00){\makebox(0,0)[cc]{\large D}}
\drawline(110,20)(125,28)\drawline(110,20)(125,12)
\put(50.00,10.00){\makebox(0,0)[cc]{\large $\varphi$}}
\end{picture}
\caption{\small Abstract outline of an interferometric detection scheme.}
\label{f:out}\end{figure}
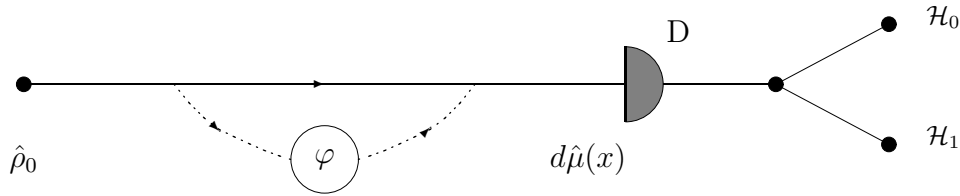\vspace{10pt}
In this paper we address the interferometric problem as a binary
decision one. In the next Section we briefly review the binary
decision problem as solved by a Neyman-Pearson optimized strategy.
In Section \ref{s:bdp} we state the binary interferometric problem.
After the illustrative example of coherent states we consider the
optimal detection scheme, according to the Neyman-Pearson criterion, for
a generic pure state of radiation. A general bound of precision is
thus obtained in terms of photon number fluctuations.
Two different classes of phase-enhanced states of radiation has been
then considered: squeezed states and phase-coherent states. The corresponding
bounds on the minimum detectable phase-shift are also evaluated.
Section \ref{s:out} closes the paper with some concluding remarks.
\section{Neyman-Pearson strategy for binary decision}
\label{s:nps}
Our goal is to determine whether or not the initial density matrix has
been perturbed. Starting from the outcomes of the detector $D$
we have to infer which is the state of the system,
in order to discriminate between the following two hypothesis:
\begin{itemize}
\item[${\cal H}_0$:] No perturbation has been occurred:
                     true if we infer $\hat\rho_0$;
\item[${\cal H}_1$:] The system has been perturbed:
                     true if we infer $\hat\rho_1$.
\end{itemize}
We denote by $P_{01}$ the probability of wrong inference for the
hypothesis ${\cal H}_1$, namely that
one of inferring ${\cal H}_1$ when ${\cal H}_0$ is true. In hypothesis
testing formulation this is usually referred to as {\it false alarm
probability}. Conversely, we denote by $P_{11}$ the {\it detection
probability}, that is the probability of inferring ${\cal H}_1$ when
it is actually true. \par
Now, which is the best measurement to discriminate between
$\hat\rho_0$ and $\hat\rho_1$ ? \par
If these two states are mutually orthogonal the problem has a trivial
solution. It is a matter of measuring the observable for which
$\hat\rho_0$ and $\hat\rho_1$ are eigenstates. However, this is
not our case, as it is well known that no orthogonal set of
phase-eigenstates is available in quantum optics.
In the following we consider nonorthogonal $\hat\rho_0$ and $\hat\rho_1$ and
we focus our attention on pure states $\hat\rho_0 = |\psi_0 \rangle\langle
\psi_0 |$ as input for the interferometer. \par
The optimization problem can be analytically solved, for pure states,
by adopting, the Neyman-Pearson criteria for binary decision \cite{nep}.
The latter reads as follows. First, we have to fix a value for the
false alarm probability $P_{01}$. Then, we have to find the measurement
strategy $d\hat\mu (x)$ which maximizes the detection probability $P_{11}$.
As a general definition, each measurement strategy which maximizes the
detection probability $P_{11}$ for a fixed value of false alarm
probability  $P_{01}$
is considered as a Neyman-Pearson optimized detection for binary
hypothesis testing. It was shown by Helstrom \cite{hel} and Holevo
\cite{hol} that this very general
problem could be reduced to solving the eigenvalue problem
for the operator
\begin{eqnarray}
d\hat\mu (x|\lambda ) = \hat\rho_1 -\lambda\hat\rho_0
\label{opt}\;,
\end{eqnarray}
which represents the optimized measurement scheme.
The parameter $\lambda$ is a Lagrange multiplier. Different
values of $\lambda$ correspond to different values of the false alarm
probability, namely to a different Neyman-Pearson strategies. \par
Once the eigenvalues problem for $d\hat\mu (x|\lambda )$ has been solved it
results that only positive eigenvectors contribute to the detection
probability $P_{11}$ \cite{hel,var}.
Thus the decision strategy is transparent:
after a measurement of the quantity $d\hat\mu (x|\lambda )$ if the outcome is
positive we infer that perturbation hypothesis ${\cal H}_1$ is true.
Conversely, we infer null hypothesis ${\cal H}_0$ when obtaining negative
outcome. By expanding the eigenstates of $d\hat\mu (x|\lambda )$ in terms of
$|\psi_0\rangle$ and $|\psi_1\rangle$ the Lagrange multiplier $\lambda$ can be
eliminated from the expression of detection probability which results
\begin{eqnarray}
P_{11} = \left\{\begin{array}{cr}
\left[\sqrt{P_{01} \kappa} + \sqrt{(1-P_{01})(1-\kappa)}\right]^2
& 0\leq P_{01} \leq \kappa \\ & \\1 & \kappa \leq P_{01} \leq 1
\end{array}\right.
\label{dprob}\;.
\end{eqnarray}
In Eq. (\ref{dprob}) $\kappa$ denotes the square modulus of the overlap
between perturbed and unperturbed state,
in formula
\begin{eqnarray}
\kappa = \left|\langle \psi_0 | \psi_1\rangle\right|^2 =
\left|\langle \psi_0 |\exp\{i\hat n\varphi\} |\psi_0\rangle\right|^2
\label{overlap}\;.
\end{eqnarray}
The overlap depends both on the initial state and on the occured
phase-shift $\varphi$. It is obvious that if the overlap is small,
it is easy to discriminate between
the two states. Thus, it is possible to obtain strategies with large
detection probability without paying the price of an also large false
alarm probability. On the contrary, if the overlap becomes appreciable it
is difficult to discriminate the states. In the limit of complete overlap
the perturbed and the unperturbed states become indistinguishable.
Detection probability is now equal to false alarm probability and
the decision strategy is just a matter of guessing after each random
measurement outcome. \par
Choosing a value for the false alarm probability is a matter of
convenience, depending on the specific problem this approach would be
applied. The maximum tolerable value for $P_{01}$ increases with the
expected number of measurement outcomes, and conversely a very low rate
detection scheme needs a very small false alarm probability.
The latter is the case of interferometry, in the following we always
will consider small value for $P_{01}$.
\section{Interferometry as a binary decision problem}
\label{s:bdp}
Once an input state for the interferometer has been specified the
probability measure in Eq. (\ref{opt}) defines the detection scheme to be
performed in order to implement an optimized interferometer.
Optimality is in the Neyman-Pearson sense, namely that detection
(\ref{opt}) maximizes the detection probability $P_{11}$ at a fixed
tolerable value of the false alarm probability $P_{01}$. \par
The interferometric strategy is thus the following: the initial state
$\hat\rho_0$ is prepared and left free to travel along the
interferometer. A set of measurements for the quantity (\ref{opt}) is
then  performed and from the data record we have to infer the state of
radiation at the output of the interferometer.
From this inference we can discriminate
in between the two hypothesis, namely we are able to monitor
the optical path of the light beam. \par
The input state is fixed in advance, therefore the
detection probability depends only on the accepted value for the false
alarm probability and on actual value of the phase-shift $\varphi$.
The minimum detectable value of the phase-shift, denoted by $\varphi_M$,
is defined by the relation
\begin{eqnarray}
P_{11} (\varphi_M ;P_{01} ) = \frac{1}{2}
\label{mdp}\:.
\end{eqnarray}
A lower value for $P_{11}$, in fact, would make the measurements record
unuseful,
as no readable informations can be extracted in that case.  \par
Let us consider customary coherent states as an illustrative example.
Without loss of generality we can set the phase of initial state to be
zero, so that we have
\begin{eqnarray}
|\psi_0\rangle = \exp\{-\frac{1}{2} \alpha^2\} \sum_{k=0}^{\infty}
\frac{\alpha^k}{\sqrt{k!}} \: |k\rangle \qquad \alpha \in {\bf R}
\label{coh}\:.
\end{eqnarray}
The photon distribution of a coherent state is Poissonian with mean
given by $N\equiv\langle\alpha |\hat n |\alpha\rangle = \alpha^2$
and the overlap can be easily evaluated to be
\begin{eqnarray}
\kappa =
\left|\langle|\psi_0| e^{i\hat n \varphi}|\psi_0\rangle\right|^2
=\exp\{-2 \alpha^2 (1-\cos\varphi )\}
\label{ovcoh}\:.
\end{eqnarray}
Interferometric detection is frequently involved with low rate
processes
\footnote{\footnotesize Among applications of high sensitive
interferometry one of the most interesting regards the detection of
gravitational waves. The reader may agree that this is a prototype for
{\em very} low rate process \cite{rmp}.}.
Therefore, we have, as a general requirement, to ask for a small false alarm
probability: a convenient setting reads $P_{01} \leq \kappa$.
Inserting Eq. (\ref{dprob}) in Eq. (\ref{mdp}) the relation for the
minimum detectable phase shift becomes
\begin{eqnarray}
\frac{1}{2} = \left[ \sqrt{P_{01} \kappa} + \sqrt{(1-P_{01})
(1-\kappa)}\right]^2
\label{min}\;,
\end{eqnarray}
that is,
\begin{eqnarray}
\kappa= \sqrt{\frac{1+\sqrt{P_{01}(1-P_{01})}}{2}}
\label{kappa}\:.
\end{eqnarray}
Finally, upon substituting (\ref{ovcoh}) in Eq. Eq. (\ref{kappa}) and
expanding for small $\varphi$ we have
\begin{eqnarray}
\varphi_M = \sqrt{\log\left( \frac{2}{1+\sqrt{P_{01}(1-P_{01})}}\right)}
\frac{1}{\sqrt{N}}
\label{mincoh}\;,
\end{eqnarray}
which represents the lower bound on minimum detectable phase-shift
for {\em any} interferometer based on coherent states. The bound in Eq.
(\ref{mincoh}) is well known and represent the lower bound on precision
also for interferometer based on classical state of radiation. It is
usually termed shot-noise limit. \par
Let us now consider a generic pure state at the input of the
interferometer
\begin{eqnarray}
|\psi_0\rangle = \sum_{k=0}^{\infty} c_k |k\rangle
\label{gen}\;.
\end{eqnarray}
Still we consider zero initial phase, thus the coefficients
$\{ c_k \}_{k \in {\bf N}}$ are real numbers. For the overlap we have
\begin{eqnarray}
\kappa = \left| \sum_{k=0}^{\infty} c_k^2 e^{i k \varphi}\right|^2
= \left( \sum_{k=0}^{\infty} c_k^2 \cos k\varphi\right)^2 +
\left(\sum_{k=0}^{\infty} c_k^2 \sin k\varphi\right)^2
\label{ovgen}\:,
\end{eqnarray}
and, up to second order in the phase-shift,
\begin{eqnarray}
\kappa = 1 - \varphi^2 \Delta N^2
\label{ovgen2}\;,
\end{eqnarray}
where $\Delta N^2 = \langle{\hat n}^2\rangle - \langle\hat n \rangle^2$
denotes the photon number fluctuations of the considered state.
By substitution in Eq. (\ref{kappa}) we obtain the lower bound on the
minimum detectable perturbation
\begin{eqnarray}
\varphi_M = \sqrt{\frac{1-\sqrt{P_{01}(1-P_{01})}}{2}}
\frac{1}{\Delta N}
\label{mingen}\;.
\end{eqnarray}
Eq. (\ref{mingen}) represents a quite general result.
It indicates that the minimum detectable phase-shift $\varphi_M$ shows
an inverse scaling relative to the photon number fluctuations rather
than the photon number intensity.
Eq. (\ref{mingen}) is not surprisingly, however it is worth
noticing that we derived it in a direct way from the binary problem approach,
i.e. we did not make use of any uncertainty phase-number
{\em pseudo} relation $\Delta N \Delta\varphi \sim 1$. The latter, in
fact, can be derived only in a heuristic way \cite{cav}
and thus possesses only a limited validity. \par
Eq. (\ref{mingen}) suggests to use
states with equally probable photon number excitation, namely
\begin{eqnarray}
|\psi\rangle = \sum_k c_k |k\rangle \qquad \hbox{with}\;\;  c_k = z \in
{\bf C}:, \;\;\: \forall k
\label{equa}\;.
\end{eqnarray}
Such a states, in fact, show infinite photon number fluctuations thus
allowing, in principle, to monitor an optical path with arbitrary
precision. Unfortunately, the only possibility to construct
states as in Eq. (\ref{equa}) is given by the London phase-states
\cite{lon,sug}
\begin{eqnarray}
|\phi\rangle = \sum_k \exp\{i k \phi\} |k\rangle
\label{SG}\;,
\end{eqnarray}
which neither possess a finite mean photon number nor they are
normalizable, namely they are not realistic states of radiation. \par
A realistic approximation of London phase-states is provided by
the so-called phase-coherent states \cite{sha}
\begin{eqnarray}
|\chi\rangle = \sqrt{1-|\chi |^2}\: \sum_k \chi^k \: |k\rangle
\label{phcoh}\;,
\end{eqnarray}
where the complex number $\chi = x\exp\{i\phi\}$ is confined in the unit
circle $x < 1$ to assure normalization. A phase-coherent coherent state
possesses a mean photon number given by $N=x^2 (1-x^2)^{-1}$
and goes to a London phase-state in the limit $x \rightarrow 1$.
For a phase-coherent state with zero initial phase the overlap
reads as
\begin{eqnarray}
\kappa = \left|(1-x^2) \sum_k x^{2k} e^{ik\varphi} \right|^2  =
\frac{1 + x^4 - 2 x^2}{1+ x^4 - 2 x^2 \cos\varphi}
\label{ovphcoh}\;,
\end{eqnarray}
leading to a lower bound on minimum detectable perturbation given by
\begin{eqnarray}
\varphi_M= \sqrt{\frac{1-\sqrt{P_{01}(1-P_{01})}}{1+\sqrt{P_{01}(1-P_{01})}}}
\frac{1}{\sqrt{N(N+1)}}
\label{minphcoh}\;.
\end{eqnarray}
The $\varphi_M$ scaling in Eq. (\ref{minphcoh}) is largely improved
relative to the shot noise limit and represents the benefit of using
phase-coherent states. \par
We now proceed by considering squeezed-coherent states at the input of
the interferometer. The benefit of squeezing in improving precision
is quite known for specific setup, as Mach-Zehnder or Michelson
interferometer\cite{yur,cao,par}. Here we are obtaining a more general bound,
which are independent on the measurement strategy. \par
We consider the interferometer fed by an in-phase squeezed-coherent
state, namely a state with parallel signal and squeezing phases. We set this
value to zero, so that the initial state is given by
\begin{eqnarray}
|\psi_0\rangle = \hat D(x) \hat S(r) |0\rangle
\label{squ}\:,
\end{eqnarray}
being $\hat D = \exp\{ \alpha a^{\dag} -\bar\alpha a\}$ and
$\hat S = \exp\{ 1/2 (\zeta a^{\dag 2} -\bar\zeta a^2)\}$ the
displacement and the squeezing operator respectively.
The parameter $\alpha$ and $\zeta$ are generally
complex, by choosing zero initial phase we can set $\alpha=x \in {\bf R}$
and $\zeta = r \in {\bf R}$. The quantity $x$ represent the coherent
amplitude of the signal whereas $r$ is the squeezing parameter. The mean
photon number of such a state is given by $N=x^2 + \sinh^2 r$. We refer
to this two terms as the signal and the squeezing photons number
respectively. \par
A perturbation in the optical path acts differently
on the coherent signal relative to the squeezing part, more precisely
we have
\begin{eqnarray}
e^{i\hat n \varphi}\; |\alpha,\zeta\rangle =
|\alpha e^{i\varphi},\zeta e^{i2\varphi} \rangle
\;.\label{evol1}
\end{eqnarray}
The overlap is thus expressed by
\begin{eqnarray}
\kappa = \left|\langle 0| \hat S^{\dag}(r) \hat D^{\dag}(x) \hat D(x
e^{i\varphi}) \hat S(r e^{i2\varphi}) |0\rangle\right|^2
\label{ovsq0}\:,
\end{eqnarray}
which is a double Gaussian curve
\begin{eqnarray}
\kappa = \frac{1}{2\sigma_1\sigma_2} \exp\left\{
-x^2\left[ \frac{(1-\cos\varphi)^2}{2\sigma_1^2}
+\frac{\sin^2 \varphi}{2\sigma_2^2}\right]]\right\}
\label{ovsq1}\:,
\end{eqnarray}
with variances given by
\begin{eqnarray}
\sigma_1 &=& \frac{1}{8}\left[e^{2r}(3+\cos 2\varphi)+e^{-2r}(1-\cos
2\varphi)\right]  \nonumber \\
\sigma_2 &=&  \frac{1}{8}\left[e^{2r}(1-\cos 2\varphi)+e^{-2r}(3+\cos
2\varphi)\right]
\label{s1s2}\:.
\end{eqnarray}
Up to second order in the phase-shift we obtain
\begin{eqnarray}
\kappa \simeq 1- \varphi^2 x^2 \sigma_1^2 =
1- \varphi^2 2\beta (1-\beta)\: N^2
\label{ovsq2}\:,
\end{eqnarray}
being $\beta$ the fraction of the total number of photons which is
engaged in squeezing, namely $\sinh^2 r=\beta N$.
The lower bound on minimum detectable phase-shift is
then obtained by substitution in Eq. (\ref{kappa}), in formula
\begin{eqnarray}
\varphi_M = \sqrt{\frac{1-\sqrt{P_{01}(1-P_{01})}}{\beta
(1-\beta )}}\frac{1}{2N}
\label{minsq}\:.
\end{eqnarray}
The proportionality constant is of the order of one as a function
of $P_{01}$, whereas the optimum value for the squeezing fraction is given
by $\beta=1/2$.
\section{Some remarks}
\label{s:out}
In this paper we have addressed interferometry as a binary decision
problem and have derived lower bounds on minimum detectable phase-shift
for some phase-enhanced states of radiation field.
We did not concern to any specific measurement device and we do not
discuss the feasibility of optimized measurement. Actually, the optimal
detection, according to the Neyman-Pearson criterion, is generally not
available at the present time. Rather, we have attempted to derive the
ultimate quantum limit on the detectable phase-shift, which depends only
on the initial quantum state at the input of the interferometer.
It is worth noticing that for squeezed states the bound in Eq. (\ref{minsq})
can actually be approximated by homodyne detection \cite{sqv,fed} or
Mach-Zehnder interferometer \cite{hol,yur,cao,par}, however only around a
fixed value for the initial phase-shift and for high efficiency of the
involved photodetectors. \par
An ultimate, state-independent, lower bound on the interferometric
precision could be obtained by a further optimization of
Eq. (\ref{ovgen}) over quantum states of radiation, provided that some
physical constraints are satisfied. Work along this line is in progress
and results will reported elsewhere.


\begin{thebibliography}{99}
\bibitem{hel} C.W.Helstrom, {\em Quantum Detection and Estimation Theory} (Academic Press, New York, 1976).
\bibitem{hol} A. S. Holevo, {\em Probabilistic and Statistical Aspect of Quantum Theory}, (North-Holland, Amsterdam, 1982).
\bibitem{yur} B. Yurke, S. L. McCall, J. R. Klauder, Phys. Rev. A {\bf 33}, 4033 (1986).
\bibitem{scu} M. O. Scully, Phys. Rev. Lett. {\bf 55}, 2802 (1985).
\bibitem{bur} M. J. Holland and K Burnett,  Phys. Rev. Lett. {\bf 71}, 1355 (1993).
\bibitem{rip} G. M. D'Ariano, M. G. A. Paris, Phys. Rev. A{\bf 49}, 3022 (1994).
\bibitem{nep} J. Neyman, E. Pearson, Proc. Camb. Phil. Soc. {\bf 29}, 492 (1933); Phil. Trans. Roy. Soc. London A{\bf 231}, 289 (1933).
\bibitem{var} A. S. Holevo, J. Multivar. Anal. {\bf 3}, 337 (1973); H. P. Yuen, R. S. Kennedy, M. Lax, IEEE Trans. Inf. Theory, IT{\bf 21}, 125 (1975).
\bibitem{rmp} K. Thorne, Rev. Mod. Phys. {\bf 52}, 285 (1980); W. W. Chou, Rev. Mod. Phys. {\bf 57}, 61 (1985).
\bibitem{cav} A. S. Lane, S. Braunstein, C. M. Caves, Phys. Rev. A{\bf 47}, 1667 (1993).
\bibitem{lon} F. London, Z. Phys. {\bf 40}, 193 (1927).
\bibitem{sug} L. Susskind, J. Glogower, Physics (N.Y.) {\bf 1}, 49 (1964).
\bibitem{sha} J. H. Shapiro, S. R. Shepard, Phys. Rev. A{\bf 43}, 3795 (1991).
\bibitem{cao} C. M. Caves, Phys. Rev. D {\bf 23}, 1693 (1981).
\bibitem{par} M. G. A. Paris, Phys. Lett. A{\bf 201}, 132 (1995).
\bibitem{sqv} M. G. A. Paris, Mod. Phys. Lett. B, {\bf 9}, 1141 (1995).
\bibitem{fed} G. M. D'Ariano, M. G. A. Paris, R. Seno, Phys. Rev. A {\bf 54}, 4495 (1996).
\end{thebibliography}
\end{document}